\documentclass{elsart}

\usepackage[dvips]{graphicx}

\begin{document}

\begin{frontmatter}



\title{An Active Gain-control System for Avalanche Photo-Diodes under 
Moderate Temperature Variations}


\author[label1]{J. Kataoka},
\corauth[corl]{Corresponding author. Tel.: +81-3-5734-2388; fax: +81-3-5734-2388}
\ead{kataoka@hp.phys.titech.ac.jp}
\author[label1]{R. Sato},
\author[label1]{T. Ikagawa},
\author[label1]{J. Kotoku},
\author[label1]{Y. Kuramoto},
\author[label1]{Y. Tsubuku},
\author[label1]{T. Saito},
\author[label1]{Y. Yatsu},
\author[label1]{N. Kawai},
\author[label2]{Y. Ishikawa} and
\author[label2]{N. Kawabata}

\address[label1]{Tokyo Institute of Technology, 2-12-1 Ookayama, Meguro, Tokyo, 152-8551, Japan}
 \address[label2]{Hamamatsu Photonics K.K., Hamamatsu, Shizuoka, Japan}

\begin{abstract}

Avalanche photodiodes (APDs) are promising light sensor for various 
fields of experimental physics. It has been argued, however, that 
variation of APD gain with temperature could be a serious problem 
preventing APDs from replacing traditional photomultiplier tubes (PMTs) 
in some applications. 
Here we develop an active gain-control system to keep the APD gain
stable under moderate temperature variations. As a performance
demonstration of the proposed system, we have tested the response of 
a scintillation photon detector consisting of a 5$\times$5 mm$^2$ 
reverse-type APD optically coupled with a CsI(Tl) crystal. 
We show that the APD gain was successfully controlled under a 
temperature variation of $\Delta T$ = 20$^{\circ}$C, within a time-cycle 
of 6000 sec. The best FWHM energy resolution of 6.1$\pm$0.2 $\%$ 
was obtained for 662 keV $\gamma$-rays, and the energy threshold was as 
low as 6.5 keV, by integrating data from +20$^{\circ}$C $-$ 0$^{\circ}$C 
cycles. The corresponding values for $-$20$^{\circ}$C $-$ 0$^{\circ}$C
 cycles were 6.9$\pm$0.2 $\%$ and 5.2 keV, respectively. 
These results are comparable, or only slightly worse than that obtained 
at a fixed temperature. Our results suggest new potential uses for 
APDs in various space researches and nuclear physics. As 
examples, we briefly introduce 
the $NeXT$ and Cute-1.7 satellite missions that will carry the APDs  
as scientific instruments for the first time.  
\end{abstract}

\end{frontmatter}

\section{introduction}

Silicon avalanche photodiodes (APD;\cite{web74} see also 
\cite{mos98,mos99,mos01,mos02,mos03,ika03,kat04}) are the only devices 
which have internal gain. Since the good features of both photodiodes (PDs) 
and photomultiplier tubes (PMTs) are shared in a single device,  
APDs offer new design for physics experiments and devices for nuclear 
medicine. In our previous works, we have studied the performance of 
reverse-type APDs{\footnote{We note that the reverse-type APD refers to a reach-through design where the narrow, high-field multiplication region is near the front of the device \cite{ika05,kat05}.}}
as an X-ray/$\gamma$-ray detector optically 
coupled with various inorganic scintillators, such as CsI(Tl), BGO,
GSO(Ce), and YAP(Ce) \cite{ika05,kat05}. 
Thanks to its high quantum efficiency (QE) and low noise contribution, 
reverse-type APDs generally show much better energy 
resolution than traditional PMTs. Moreover, due to their  
internal gain of 50 or more, we have successfully used these APDs to detect  
weak scintillation light from CsI(Tl) crystal at energies as low as 5 keV even 
at room temperature. Most recently, we have successfully developed 
various large area reverse-type APDs with Hamamatsu Photonics, 
up to 20$\times$20 mm$^2$ square area\cite{sat06}. 
This is the one of the largest monolithic APD pixel ever 
produced.
Such large dimensions have been awaited by researchers world-wide,  
and further extend the potential of APDs for various 
applications in the near future. 

It should be noted, however, that the gain characteristics of APDs 
depend on both the bias voltage and the temperature. Although the 
gain variation of APDs on bias voltage ($\sim$ 3$\%$/V at a 
gain of 50) is only two times larger than the voltage coefficient of typical 
PMTs, the temperature variation can be a more critical problem for 
APDs. When the APD device is cooled, the bias voltage required to 
achieve a certain gain is significantly reduced. 
This is due to the much smaller probability of electron 
energy loss in interactions 
with phonons as compared with that at room temperature. In fact, 
the temperature coefficient of APDs ($\sim$ $-$2$\%$/$^{\circ}$C) is 
about ten times larger than that of typical PMTs\cite{ika03,ika05}. 
Therefore, in order to stabilize the gain at the 1 $\%$ level,  
the temperature must be 
controlled within $\Delta T$ $\simeq$ 0.5$^{\circ}$C, which is often 
too severe a requirement for detectors used in space missions, 
and for the large calorimeters used in the accelerator experiments. 
It should be noted that the temperature variation is $not$ 
very rapid in most of the cases discussed above. For example, 
temperatures of detectors in satellites or balloon experiments  
can vary by a large amount, but typically  $\le$ 20$^{\circ}$/hour (e.g., 
http://lss.mes.titech.ac.jp/ssp/cubesat/index\_e.html for the Tokyo Tech
pico-satellite CUTE-I).  Such requirements are generally much less stringent 
for ground-based experiments.

In order to stabilize the gain of an APD detector against temperature 
variations, a number of authors have proposed different ideas to meet their 
specific purposes. For example, it has been argued that the temperature 
compensatation should keep the APD performance near the best multiplication 
factor defined as the highest signal-to-noise ratio (\cite{dia98}, 
see also \cite{cas84}). This is a very simple design using a 
feedback loop of analog electronics, where an appropriate bias 
voltage is automatically generated at different temperatures. An obvious 
weakness is that the bias is corrected only linearly against the
temperature ($\Delta$$V$ $\propto$ $\Delta$$T$), 
which is not generally true in usual applications 
(see also $\S$2.1 and Figure 4($left$)). Such a non-linearity was carefully 
considered by different authors~\cite{pro03} but required careful 
tuning of circuit parameters used in the gain control system,
a time-consuming process, especially when the responses of 
individual APDs can differ significantly. These authors have
mentioned that they are currently developing a more convenient system 
using a CPU, where the temperature compensation can be easily done 
by a software program. Apparently, such a CPU-based system is much 
more flexible and makes it easier to include any irregularities and 
non-linearities in the APD gain characteristics.

In this paper, we present a novel, CPU-based design to keep 
the APD gain stable under moderate temperature variations. 
The paper is organized as follows. In $\S$2, we show the 
gain characteristics of an APD to explain the 
basic concept of active gain-control system. As a performance
demonstration of the proposed system, we have made a scintillation 
photon detector consisting of a 5$\times$5 mm$^2$ reverse-type APD 
optically coupled with a CsI(Tl) crystal. In $\S$3, we demonstrate that
our system works well under moderate temperature variations 
between $-$20$^{\circ}$C and 0$^{\circ}$C, or +20$^{\circ}$C and
0$^{\circ}$C, in a 6000 sec time-cycle. In $\S$4, we briefly 
introduce the applications of our system for the future space missions 
$NeXT$ and Cute-1.7, both of which will carry APDs as a 
scientific instrument. We also comment on the further potential of using 
an active gain-control system in ground based experiments.  
Our results are summarized in $\S$ 5.

\section{Development of Active Gain-Control System}
\subsection{APD Gain Characteristics}

We used a reverse-type APD of 5$\times$5 mm$^2$ surface area,
recently developed by Hamamatsu Photonics K.K.\ 
(Hamamatsu S8664-55). Full details of the detector performance and 
the internal structure of S8664-55 are given in the literature
\cite{ika03,kat05,sat06}. Summarizing the parameters of the APD 
assembled in this experiment, the leakage current is 1.52 nA and the 
detector capacitance is 85 pF when fully depleting the device. 
An avalanche gain of 50 is achieved for a bias voltage of 370\,V and the 
breakdown takes place at 411\,V. All these parameters were measured at 
room temperature (+25 $^{\circ}$C). Here we particularly focus on the 
APD gain characteristic to explain the basic concept of the $active$
gain-control system described in detail in following sections. 

The APD gain, $G$, can be measured under constant illumination
of a monochromatic light source by recording the photocurrent of the APD
as a function of bias voltage. We use a light-emitting diode (LED)
emitting at 525$\pm$5 nm. This wavelength is
particularly important to mimic the scintillation light from CsI(Tl) 
crystal (550 nm) as we will see in Section~3. At voltages lower than 50\,V, 
the APD gain can be regarded as unity. Figure~1 ($left$) shows
variations of APD gain as a function of bias voltage for different 
temperatures (measured from $-$20$^{\circ}$C to +20$^{\circ}$C).
At +20$^{\circ}$C, the gain reaches 30 at 353\,V, whereas only 320\,V is 
necessary to obtain the same gain in an environment cooled to 
$-$20$^{\circ}$C. Such a characteristic is more clearly seen in 
Figure~1 ($right$), where the APD gains for fixed bias voltages are 
plotted as a function of temperature. At a gain of $G$=30 and a
temperature of $T$=0$^{\circ}$C, the gain variation is approximated by 
\begin{equation}
\frac{1}{M}\frac{dM}{dT} \simeq -2.0 \%/^{\circ}C,  
\end{equation}
which is quite large compared to that of usual PMTs, but is the 
comparable to reverse-type APDs reported in the literature\cite{ika03}.
Under the conditions $G$=30 and $T$=0$^{\circ}$C, the dependence of
gain on bias voltage can be approximated by 
\begin{equation}
\frac{1}{M}\frac{dM}{dV} \simeq +2.5 \%/V.  
\end{equation}
Note, this is a $conventional$ relation assuming an
exponential change of APD gain with bias voltage.  
It can be seen that
this is an over-simplified model as there is a clear deviation 
from a straight-line in the semi-log plot of Figure~1 ($left$). 
Nevertheless, this provides a convenient way of comparing the 
APD performance with those of others 
in the literature \cite{ika03,kat04,ika05}.

An important remark on Figure~1 ($right$) is that, for a fixed APD gain 
$G$ (e.g., dashed line shows $G$=30 line),  there exists 
a one-to-one relation between the bias voltage and the 
temperature. In other words, we can $uniquely$
determine the bias voltage that produces $G$ for an arbitrary 
temperature. This is a $key$  $idea$ for the gain-control system 
discussed below. If the temperature increase by $\Delta$$T$, then the 
APD gain should  decrease by $-$$\Delta G$ when the bias voltage is fixed to
a constant value. In order to restore the original APD gain, we can 
simply increase the bias voltage, by $\Delta$V,  to cancel 
out the gain reduction. From equation (1) and (2), $\Delta$$V$ and 
$\Delta$$T$ are related by $\Delta$$V$ = $C(T,G)$$\times$$\Delta$$T$, 
where $C(T,G)$ is a constant but depends on the temperature and
the APD gain. 

\begin{figure}[hc]
\begin{center}
\includegraphics[height=6.95cm,keepaspectratio, angle=90]{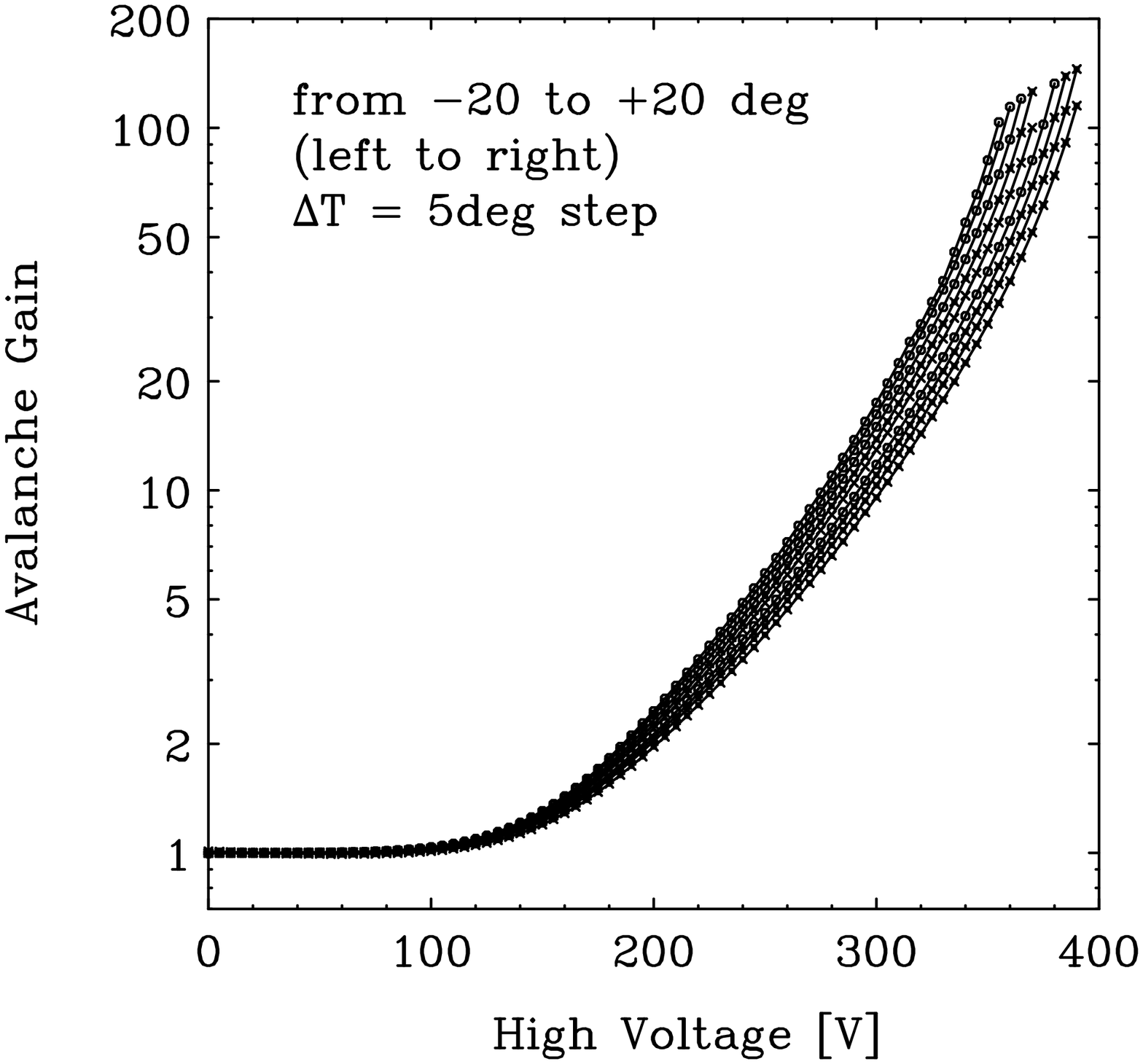}
\includegraphics[height=6.7cm,keepaspectratio, angle=90]{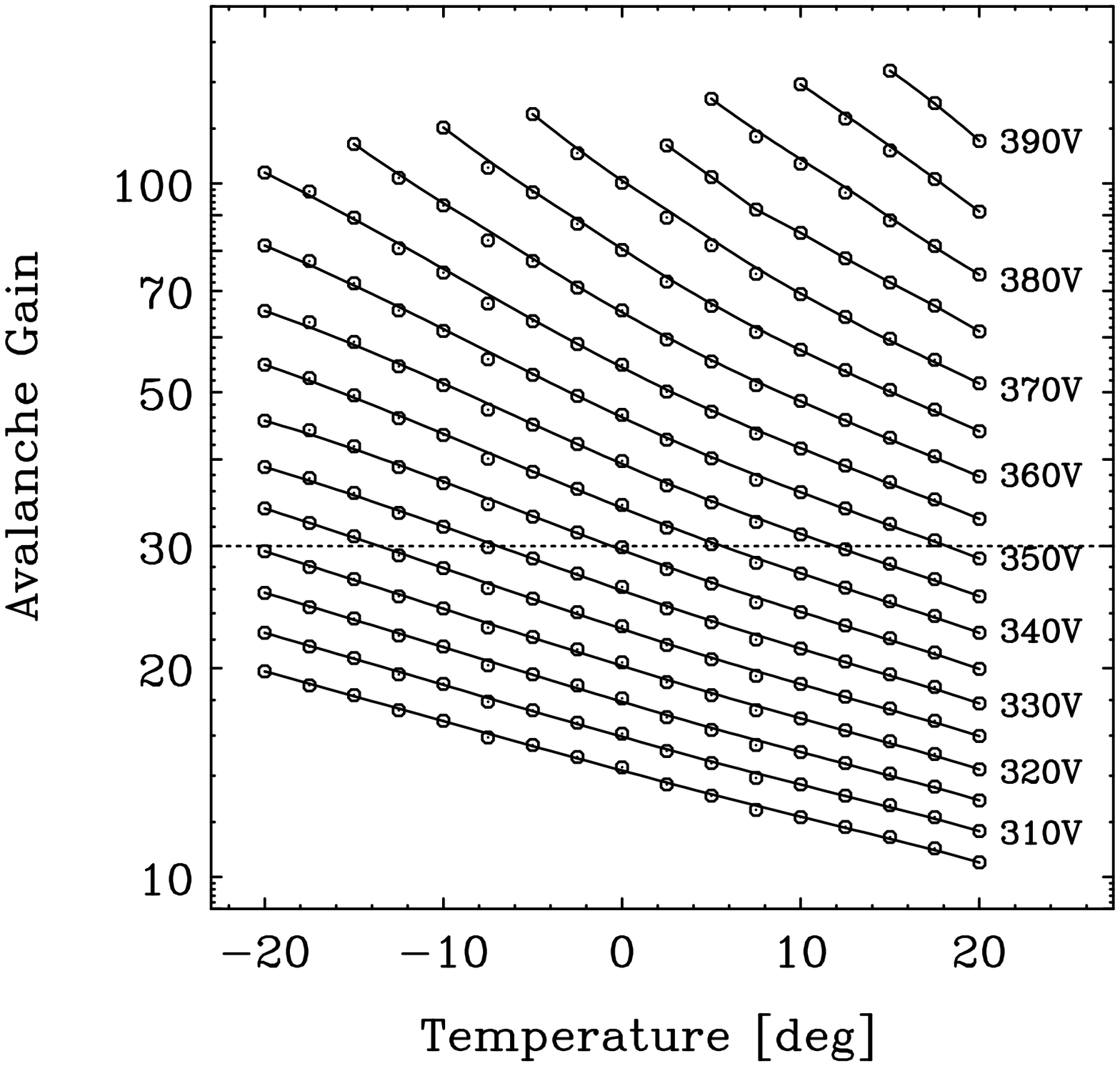}
\caption{Gain variations of APD as a function of bias voltage ($left$) 
and temperature ($right$), measured from $-$20$^{\circ}$C to  
+20$^{\circ}$C.}
\end{center}
\end{figure}

\subsection{System Design}

As a performance demonstration of the proposed idea, we made a 
scintillation photon detector consisting of a reverse-type APD 
(S8664-55; see above) optically coupled with a cubic CsI(Tl) crystal 
of dimensions 5$\times$5$\times$5mm$^3$. Figure~2 shows a picture of the 
``sensor box'', where the APD scintillation detector,  
a DC/DC converter (Analog Modules; Model 521-5) and a temperature sensor 
(Analog Device: AD590) are all implemented in a single metal box.
The DC/DC converter outputs a 0$-$600\,V bias voltage to the APD, 
by a factor of 120 amplification of the input DC signal 
(from 0 to 5\,V maximum). The AD590 is a 
2-terminal circuit temperature transducer that produces
an output current proportional to absolute temperature  (1 $\mu$A/K). 
The separation between the APD sensor (APD+CsI(Tl)) and AD590 is about 1.5 cm. 

The method of gain-control is straightforward. First, 
one should define a reference table or empirical function of 
bias voltage as a function of temperature to 
stabilize the detector gain. 
Second, the detector temperature should be monitored with an appropriate  
time interval, $\Delta t$.  
For example, if the temperature variation is expected  
on hour scales, it is recommended to monitor AD590 output every minute, or 
on even shorter time scales. 
Third, the optimum voltage to keep the detector 
gain constant is calculated, and the APD bias changed if necessary. 
Automatic 
iteration of these processes every $\Delta t$ results in a robust 
gain-control system under the moderate temperature variations.

The block diagram of the system developed in this paper 
is shown in Figure~3. The signals from the APD and the temperature
sensor were read by a peak-hold 12-bit ADC (Clearpulse; CP1113A)
connected to a Linux computer via a VME bus. 
The output signals from the APD were amplified by a charge sensitive 
preamplifier (Clearpulse; CP581K) and fed to the shaping amplifiers 
(ORTEC; 570 and 572), where in the latter, fast shaping is used to trigger 
the data acquisition. The DC output of AD590 was monitored every 1 sec, 
and then appropriate bias voltage is calculated as a function of 
temperature, in the DAQ program. This bias voltage is output from 
the VME
DIO (Clearpulse; CP2412) as 8-bit digital data, and fed to the DAC 
(Analog Device: AD558) input. The maximum output of the DAC is allowed in the 
range 2.55$-$10 V. Here we set the maximum voltage to 3.33 V, in order 
that the maximum of DC/DC converter (Analog Modules; 521-5M) becomes    
3.33 V $\times$ 120 = 400V.  Therefore a 1 bit input to the DAC corresponds to 
13.0mV$\times$120 = 1.56 V when output from the DC/DC converter. Note 
this may produce $\simeq$ 3.9 $\%$ fluctuations in APD gain if $G$ = 30 
(see $\S$2.1). We will revisit this problem in $\S$4.2
\begin{figure}[hc]
\begin{center}
\includegraphics[height=6cm,keepaspectratio, angle=0]{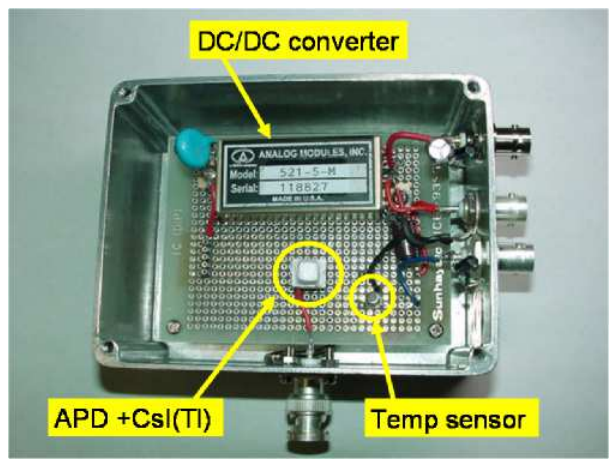}
\caption{A picture of the sensor box used in this experiment. The 
APD+CsI(Tl) scintillation detector, DC/DC converter and temperature 
sensor are all implemented in a metal box.}
\end{center}
\end{figure}

\begin{figure}[hc]
\begin{center}
\includegraphics[height=10cm,keepaspectratio, angle=0]{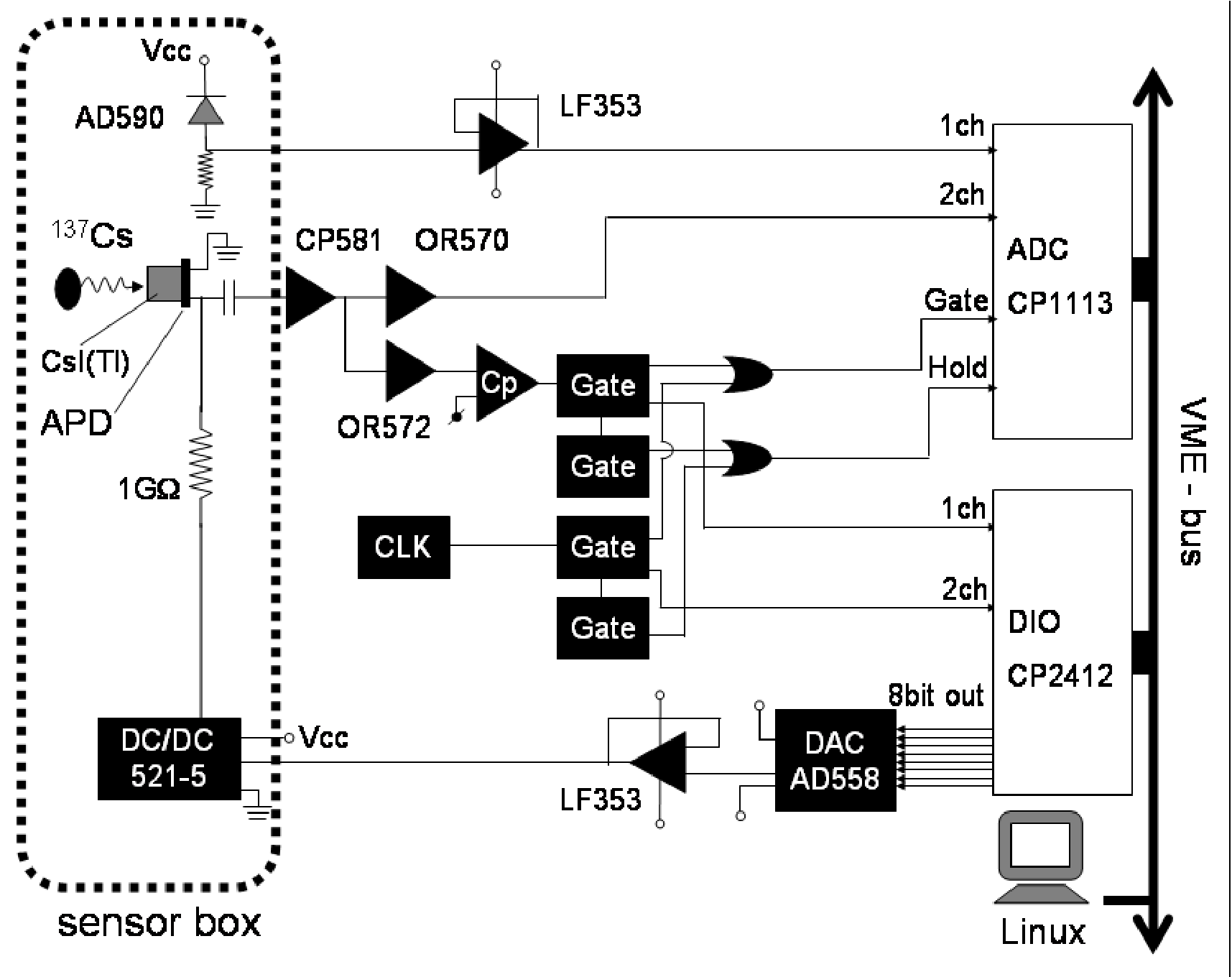}
\caption{A block-diagram of the gain-control system developed 
in this paper. A dashed box is the sensor part shown in Figure 2.}
\end{center}
\end{figure}

\section{Performance Test}
\subsection{Calibration}

Prior to the performance test, we first calibrated the APD + CsI(Tl) 
system to obtain an empirical relation between the bias voltage and 
the temperature. As a reference, we measured a spectrum of $^{137}$Cs 
$\gamma$-rays at $-$20$^{\circ}$C. The corresponding APD gain is $G$=30. 
We then calibrated the system by measuring the peak of 662 keV 
$\gamma$-rays every 2$^{\circ}$C from $-$20$^{\circ}$C to +20$^{\circ}$C. 
Actually, we only measured the bias voltage that makes the 662 keV 
peak ADC channel equal to that obtained for $-$20$^{\circ}$C, under 
various temperatures.  Note, in this system, we do not measure the 
shift of APD gain itself as described in Figure 1 ($right$), but the 
light yield of a CsI(Tl) crystal read through the APD. 
A great merit of this method is that we can 
correct not only the 
variations of APD gain, but the response of the CsI(Tl) at the same time.
In fact, it is known that the light yield of CsI(Tl) scintillator 
measured at $-$20$^{\circ}$C decreases by 15$\%$ from that measured in 
$+$20$^{\circ}$C (e.g., 
http://www.scionixusa.com/pages/navbar/scin\_crystals.html). 
Although a wide variety exists in response of
various scintillators, our approach is always applicable in correcting the 
APD gain $coupled$ $with$ the scintillator. Figure~4 ($left$) shows the
required bias voltage thus obtained, to keep the gain of scintillation 
detector as a whole, under various temperatures.  Figure~4 ($right$) 
shows the AD590 output (ADC channel; 2.5mV/ch) as a function of
temperature. From these two figures, we obtained an empirical, 
approximate relation between the bias voltage ($HV[V]$) and the 
AD590 output ($X[ch]$) as;
\begin{equation}
HV [V] = 2.3823 \times 10^{-5} {X[ch]}^2 - 8.7626 \times 10^{-2} X[ch] + 376.08\end{equation}
This $HV [V]$ is further converted to the input channel of DAC (AD558) 
via the relation
\begin{equation}
DAC_{in} [ch] = int(\frac{HV[V]-0.74482}{1.5947}), 
\end{equation}
where we only take the integral part of the right-hand side in equation (4).
Therefore, by measuring the ADC channel of temperature sensor $X[ch]$, 
we can uniquely define an appropriate bias voltage to keep the detector gain 
constant every one second.

\begin{figure}[hc]
\begin{center}
\includegraphics[height=6.8cm,keepaspectratio, angle=90]{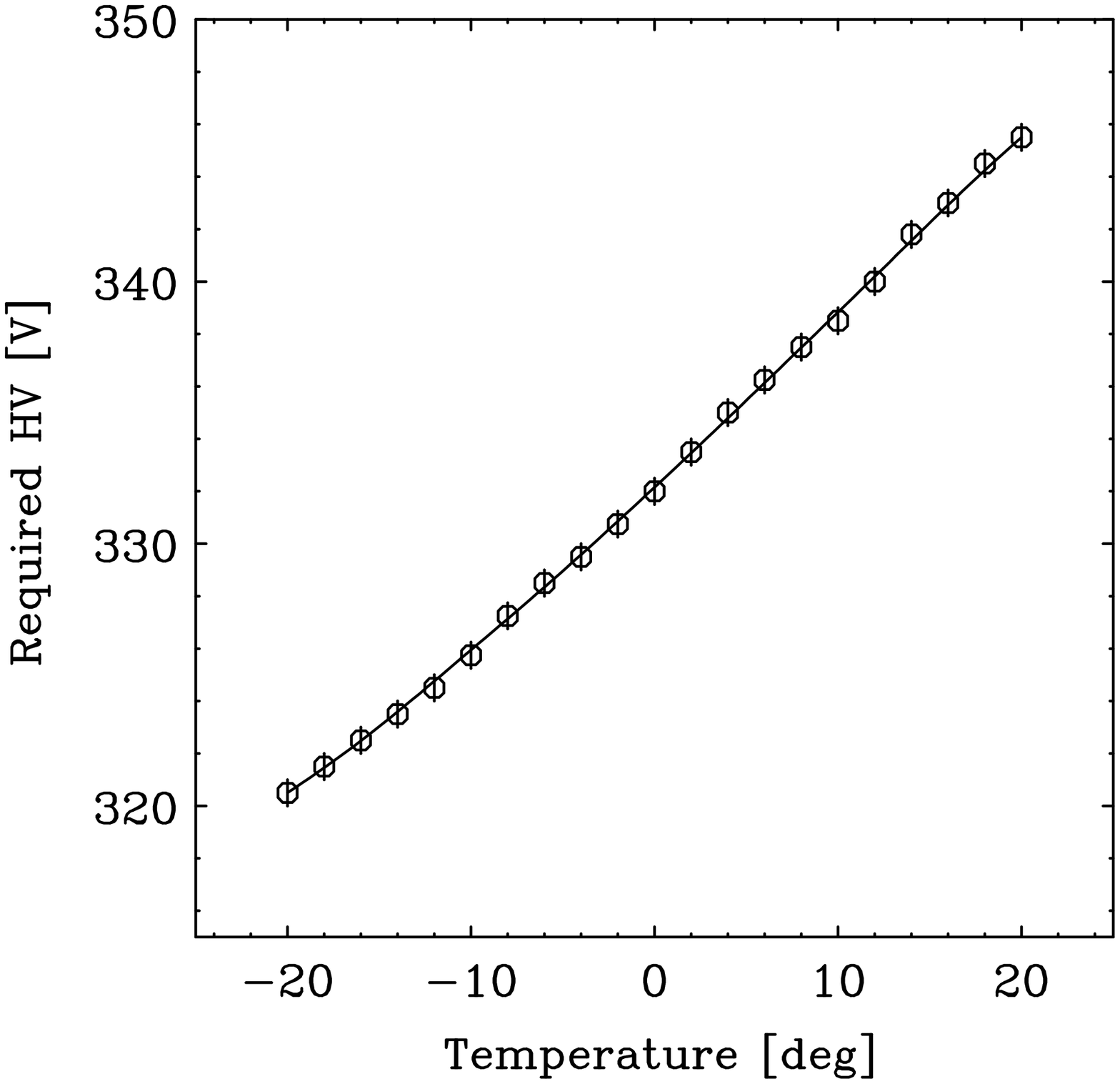}
\includegraphics[height=6.8cm,keepaspectratio, angle=90]{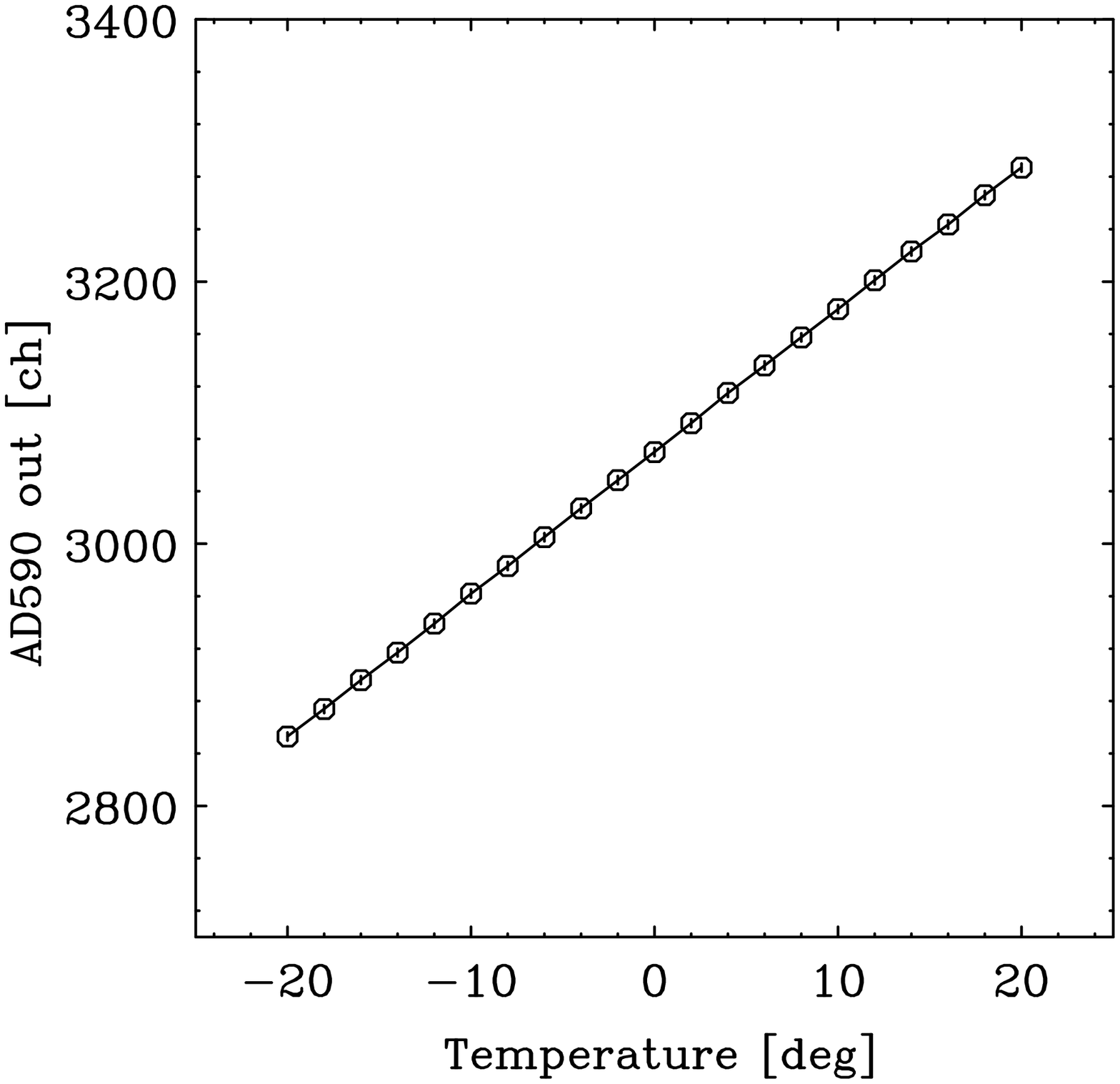}
\caption{$left$: Required bias voltage of APD + CsI(Tl) detector system 
to keep a constant gain.  The reference gain is that obtained for 
662 keV $\gamma$-rays measured at $-$20$^{\circ}$C.
$right$: Relation between the AD590 output and the temperature.
}
\end{center}
\end{figure}

\subsection{Active Gain-Control in  +20$^{\circ}$C $\leftrightarrow$
0$^{\circ}$C Cycles}

As a performance demonstration of an active gain-control system, we
measured the 662 keV $\gamma$-ray spectrum under moderate temperature 
variations. The temperature was controlled with a thermostat to vary 
between +20$^{\circ}$C and 0$^{\circ}$C, in a time-cycle of 6000 sec. 
This time constant and the temperature
range is particularly selected from the observation and the simulation 
results for the small cubic satellite missions in low-Earth orbit 
(CUTE-1 and Cute-1.7; see $\S$ 4.1 and \cite{kot06}).  
Figure~5 ($left$) shows a 
time history of  temperature variations in the thermostat 
(two cycles; thin dashes), overlaid with the measured temperature
variation reconstructed from AD590 data (thick line). 
Due to the heat capacitance of the sensor box (Figure 2), the
temperature variations of APD lags behind that in the thermostat for 
600$-$800 sec. Also, the minimum and the maximum temperature during the 
cycle is +4$^{\circ}$C and +16$^{\circ}$C, except for the very first phase 
of temperature cycle starting from +20$^{\circ}$C.

\begin{figure}[hc]
\begin{center}
\includegraphics[height=6.7cm,keepaspectratio, angle=90]{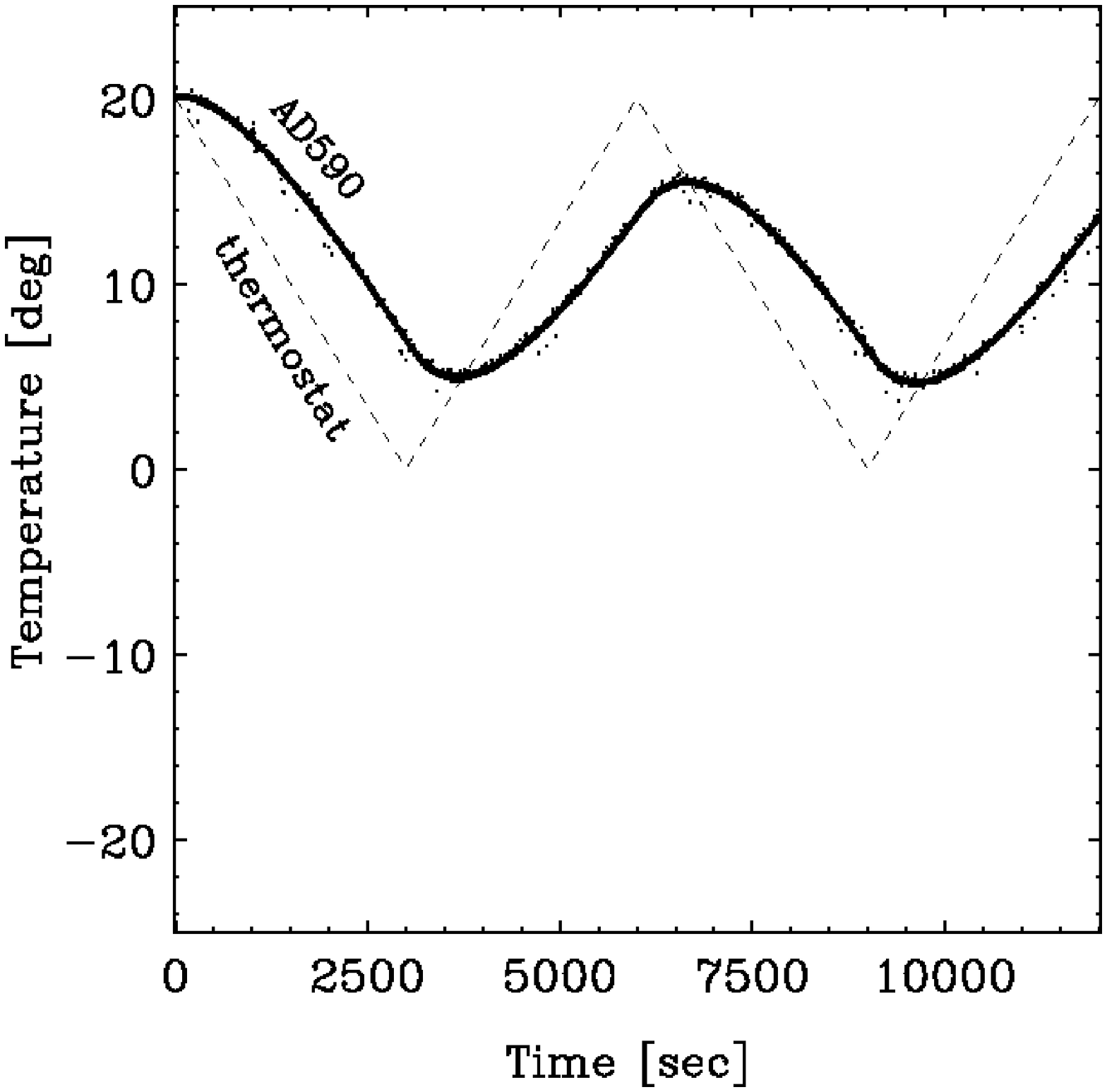}
\includegraphics[height=7.0cm,keepaspectratio, angle=90]{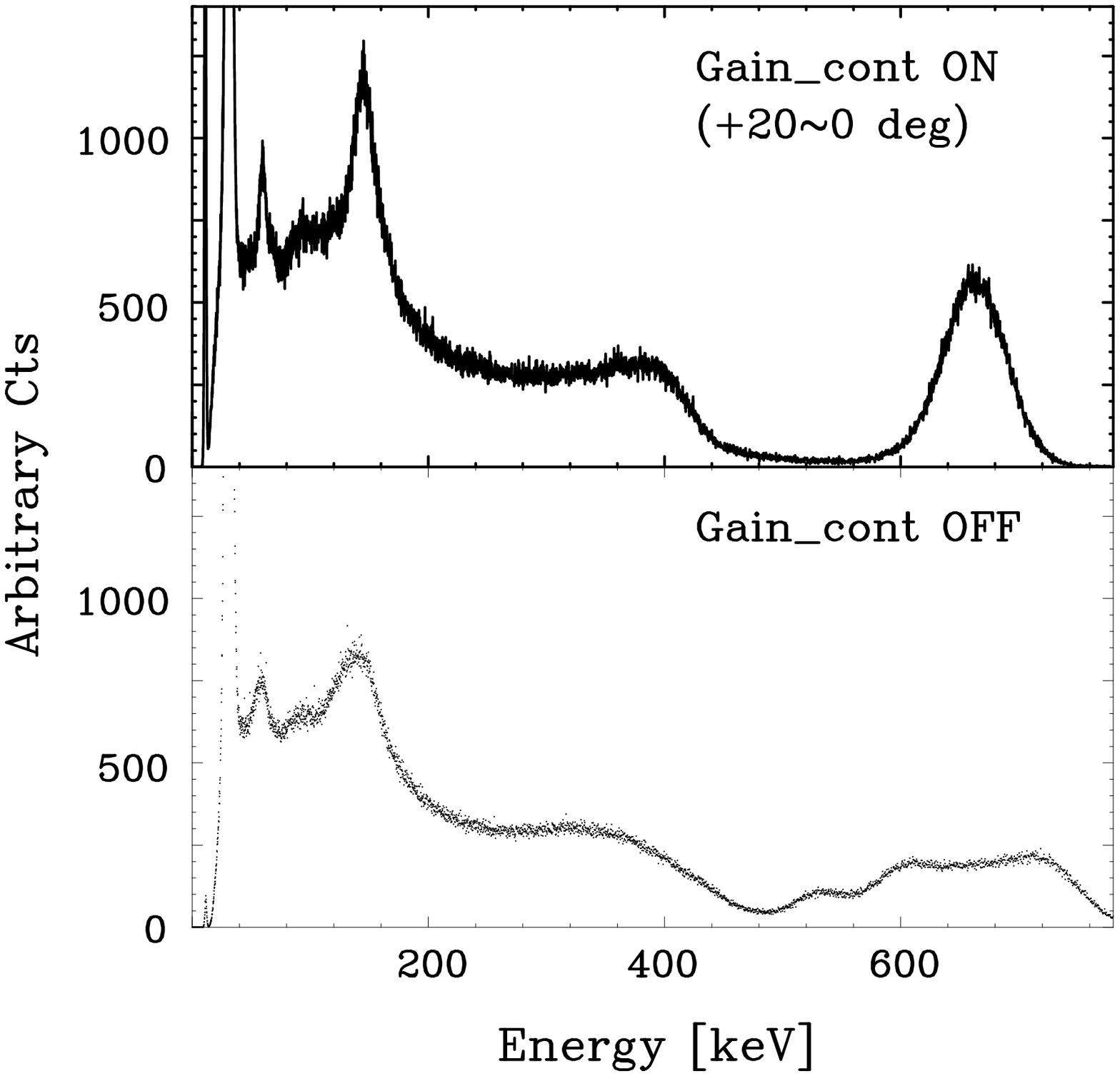}
\caption{$left$: Time history of  temperature variations 
in a thermostat (two cycles of +20$^{\circ}$C $\leftrightarrow$
 0$^{\circ}$C; thin dashed line), overlaid with the measured temperature 
of the ``APD sensor box'' reconstructed from AD590 data (thick line). 
$right$: Comparison of a 662 keV $\gamma$-ray spectra when 
the active gain control is working ($upper$) and not in use ($lower$). For the 
upper panel, the energy resolution is 6.1 $\%$ (FWHM, @662keV) and the energy 
threshold is 6.5 keV, respectively (see also Table.1).}
\end{center}
\end{figure}

Figure 5 ($right$ upper) shows a  662 keV $\gamma$-ray spectrum
integrated over two temperature cycles (12000 sec, Figure 5$ left$) 
measured with the active gain-control system. The energy resolution and 
the threshold are 6.1 $\%$ and 6.5 keV, respectively. Note that, 
these values are comparable with those 
measured at a $fixed$ temperature. For example, the corresponding energy 
resolutions  are  5.6 $\%$ (+20$^{\circ}$C) and 5.7 $\%$ (0$^{\circ}$C) 
and the energy thresholds are 7.1 keV (+20$^{\circ}$C) and 6.1 keV
(0$^{\circ}$C), respectively. To demonstrate the merits of our 
system more clearly, we also measured a 662 keV $\gamma$-ray spectrum 
integrated over the same temperature cycles,  $without$ the active 
gain-control (with the bias voltage is fixed to 332.5V). 
As shown in Figure~6 ($right$; bottom), the spectrum 
is significantly distorted, so much so that the 662 keV peak has become
a plateau rather than having a single Gaussian form. Therefore we confirm that 
the active gain-control system described in this paper provides effective 
feedback to stabilize the APD gain, at least under moderate temperature 
variations.

\subsection{Active Gain-Control in $-$20$^{\circ}$C $\leftrightarrow$ 0$^{\circ}$C  Cycles}

Next we measured the 662 keV $\gamma$-ray spectrum in a 
lower temperature cycle, between $-$20$^{\circ}$C and 0$^{\circ}$C.
As for the previous section, the temperature was controlled with a thermostat 
and slowly varied on a time scale of 6000 sec (time period for one
cycle). Figure~6 ($left$) shows a 
time history of temperature cycles in the thermostat 
(two cycles; thin dashes), overlaid with a measured temperature
variation reconstructed from AD590 data (thick line). Again, 
the temperature variation of the APD lags behind that in the thermostat by
600$-$800 sec. The actual minimum and the maximum temperature of sensor 
during the cycle is $-$16$^{\circ}$C and $-$4$^{\circ}$C, except for 
the very first phase of temperature cycle starting from 0$^{\circ}$C.

\begin{figure}[hc]
\begin{center}
\includegraphics[height=6.7cm,keepaspectratio, angle=90]{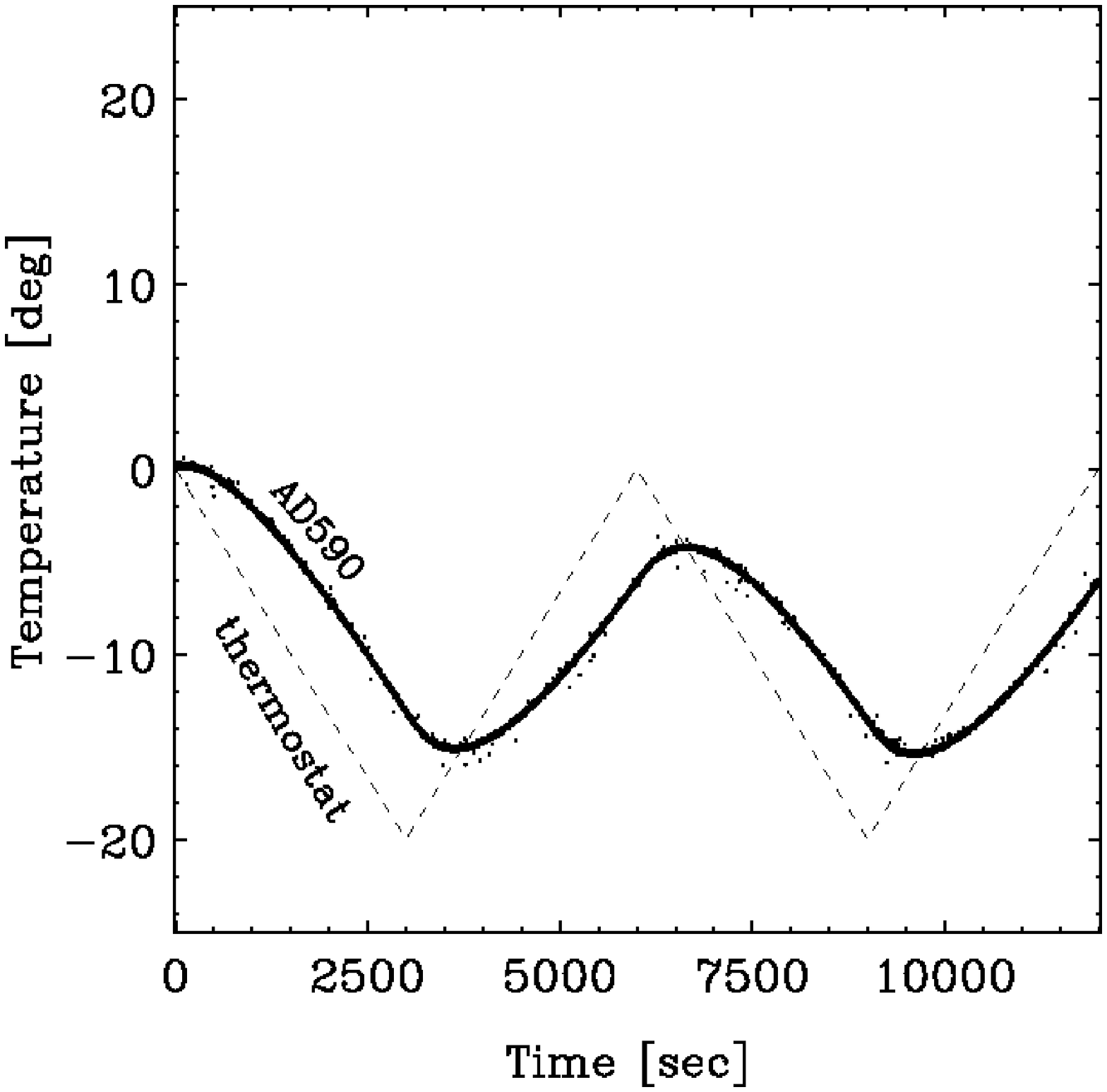}
\includegraphics[height=7.0cm,keepaspectratio, angle=90]{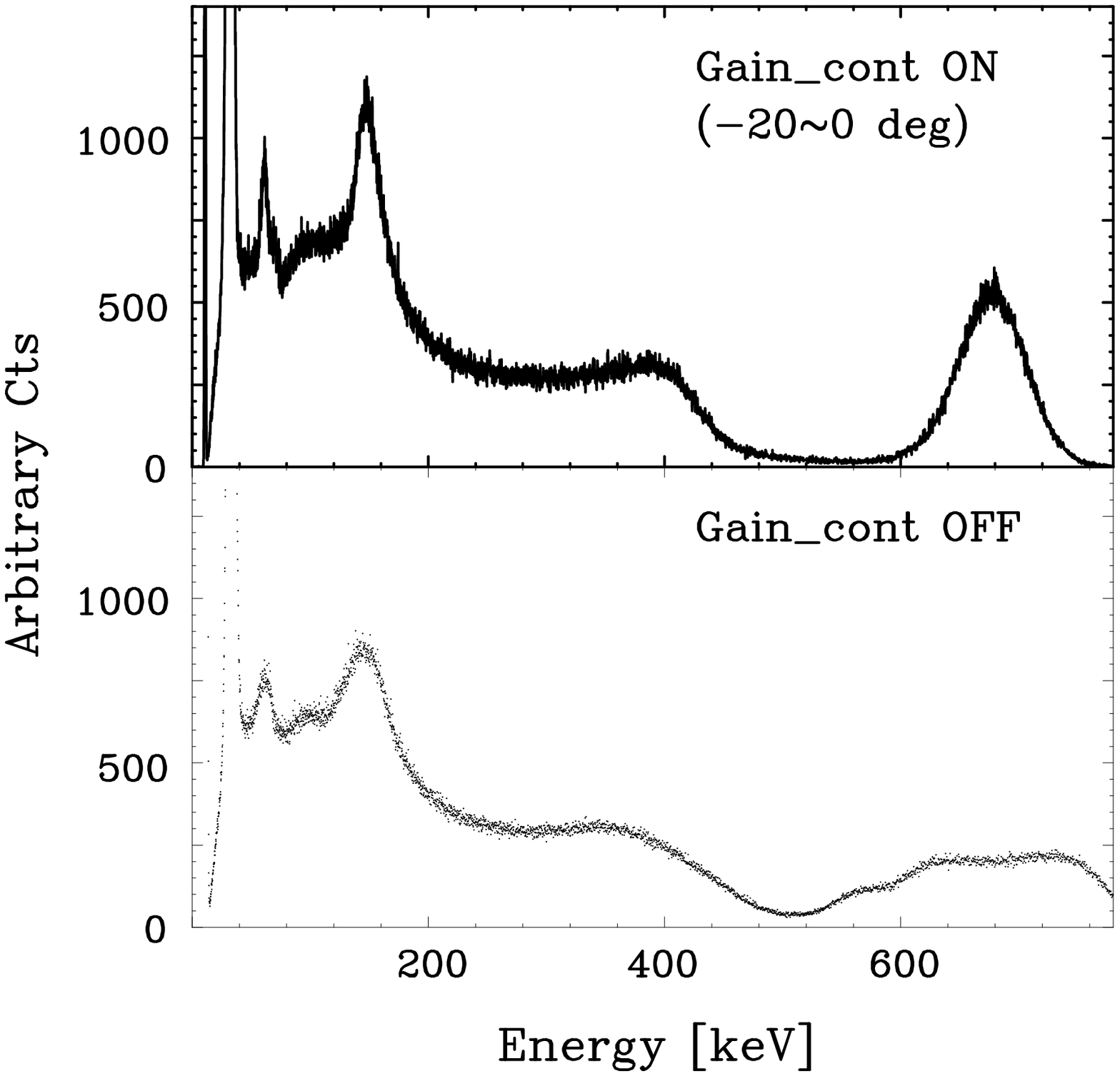}
\caption{$left$: Time history of  temperature variations 
in a thermostat (two cycles of $-$20$^{\circ}$C $\leftrightarrow$
 0$^{\circ}$C; thin dashed line), overlaid with the measured temperature 
of ``APD sensor box'' reconstructed from AD590 data (thick line). 
$right$: Comparison of a 662 keV $\gamma$-ray spectra when the automatic
 gain-control is working ($upper$) and not in use ($lower$). 
For the upper panel, the energy resolution is 6.9 $\%$ (FWHM, @662keV)
 and the energy threshold is 5.2 keV, respectively (see also Table.1).}
\end{center}
\end{figure}

Figure 6 ($right$ upper) shows a  662 keV $\gamma$-ray spectrum
integrated over two temperature cycles (12000 sec, Figure 6$ left$) 
measured with the active gain-control system. The energy resolution and 
the threshold are 6.9 $\%$ and 5.2 keV, respectively. For comparison, 
the corresponding energy resolutions  are  6.3 $\%$ ($-$20$^{\circ}$C) 
and 5.7 $\%$ (0$^{\circ}$C) and energy thresholds are 
5.1 keV (+20$^{\circ}$C) and 6.1 keV (0$^{\circ}$C), respectively, for 
fixed temperatures. 
Similar to Figure~5 ($right$), we 
compared a 662 keV $\gamma$-ray spectrum integrated over the 
same temperature cycles,  $without$ the active gain-control 
(with the bias voltage is fixed to 320.0\,V). Again we can see 
in Figure 6 ($right$; bottom) that the spectrum is significantly 
distorted. Such gain fluctuations are successfully reduced with the use of
the active gain-control system. 
 
From Table 1, we see that the energy threshold decreases with
temperature, as has been discussed elsewhere~\cite{ika03}. The sources of 
dark noise in an APD are divided into ``dark current noise'' and 
``capacitance noise'', where the former is predominant for 
S8664-55 (e.g., \cite{sat06}). Therefore significant reduction of dark 
current at lower temperature (\cite{ika03}) improves the energy 
threshold as given in Table~1. Meanwhile, the energy resolution 
is worst at $-$20 $^{\circ}$C. As briefly 
notified in $\S$3.1, the light yield of CsI(Tl) scintillator 
decreases with temperature. Assuming the energy resolution of CsI(Tl) 
is related to the incident $\gamma$-ray energy as $\propto$
$E^{-1/2}$, one should expect a $\sim$ 10$\%$ broadening in the 
energy resolution. This is indeed what we see in Table~1, 
but of curse, more careful consideration including the intrinsic 
resolution of scintillation crystal~\cite{mos02} 
is necessary for further clarification.

\begin{table}
\begin{center}
\caption{Energy Resolution and Threshold for Various Measurements.}
\begin{tabular}{r|cc|c}
\hline
Temperature & Condition & Energy Resolution & Energy Threshold\\
   & & (FWHM, @662keV) & \\
\hline
+20 $\leftrightarrow$ 0 $^{\circ}$C  &  2-cycles  & 6.1 $\%$  & 6.5 keV\\
$-$20 $\leftrightarrow$ 0 $^{\circ}$C  &  2-cycles  & 6.9 $\%$  & 5.2 keV\\
\hline
+20$^{\circ}$C  &  fixed  & 5.6 $\%$  & 7.1 keV\\
0$^{\circ}$C  &  fixed  & 5.7 $\%$  & 6.1 keV\\
$-$20$^{\circ}$C  &  fixed  & 6.3 $\%$  & 5.1 keV\\
\hline
\end{tabular}
\end{center}
\end{table}

\section{Future Plans}
\subsection{Applications in Space Satellite Missions}

Let us briefly review the advantages of using APDs.
APDs are very compact and have rugged structures. They can be operated 
under relatively low bias voltages, and are less affected by magnetic
fields. Moreover, the internal gain and high QE are great merits in 
detecting weak scintillation light, as well as direct detections of 
soft X-rays and low energy charged particles \cite{kat05}. We have a plan to 
use APDs for the X-ray/$\gamma$-ray observation satellite $NeXT$, 
which is a planned Japanese satellite which could be launched as early as 
2011\cite{tak04,tak05}. Although detailed design parameters are still 
being discussed, $NeXT$ 
will have a height of 13\,m, a diameter of 2.2\,m,
and a  total mass of 1800\,kg.
In this mission, reverse-type APDs will be implemented as a 
light sensor for large BGO plates and blocks, that effectively reject 
background cosmic rays and atmospheric $\gamma$-rays \cite{ika05}. 
A typical orbit for Japanese X-ray missions is a low-Earth, circular
trajectory with an altitude of 500$-$600 km. The inclination and 
the orbit cycle would be 31$^{\circ}$ and $\sim$6000 sec,
respectively. In such orbits, the detector temperature is easily
controlled within a range of  $-$15$\pm$5 $^{\circ}$C, without the aid of
an additional heater or thermostat. Therefore the expected temperature 
variation is quite similar to the range we have tested in $\S$ 3.3.

It is, however, noted that APDs have not previously been used in the space 
environment as a radiation detector. In general, semiconductor 
devices experience a degeneration of performance as the total radiation dose 
increases. We have therefore conducted a proton beam test to
carefully estimate the radiation effects caused by high energy 
charged particles in orbit, and found that reverse-type APDs are 
sufficiently tolerant to be used in low-Earth orbit\cite{kot05}.
For further qualification tests, we will carry the same reverse-type APD 
onboard the forthcoming pico-satellite Cute-1.7. 
We will examine the active gain-control system on-board using 
a personal digital assistant (PDA) for the first time in space \cite{iai04}.  
A detailed report for the expected APD performance in orbit is 
given elsewhere \cite{kot06}.



\subsection{Further Comments on Active Gain-Control System}

The original motivation of developing the gain-control system 
was to use the APD in satellite missions, where significant
temperature variation is inevitable.  However, this same system can be 
of course useful in various ground-based experiments. For example, 
in the field of particle physics, it is quite difficult to maintain a 
whole detector system (e.g., large scintillation calorimeters
surrounding an accelerator) at the same temperature. 
Fortunately, in most cases, temperature variation is not as large
as in a space environment, and variability occurs on relatively 
longer time scales (e.g., a few hours to days).
Therefore, the monitoring the temperatures at various 
detector positions will enable the APD gain to be kept at a constant 
level.

Meanwhile, we should also keep in mind that the active gain-control 
system may not work properly if (1) the thermistors do not follow 
the $actual$ APD temperature, and 
(2) variation time scale is too short compared to time interval 
on which the bias voltage is changed. Moreover,  
``digitization error'' may affect 
the accuracy of the gain reconstruction.  
For example, in our experimental system, the 12-bit ADC converts the output of 
thermistor with an accuracy of 0.092$^{\circ}$C/ch (Figure 4 $right$).
This will make the gain fluctuation of only $\sim$ 0.2$\%$ level 
(Equation.(1)) and will not be a serious matter. 
However, when setting an 
appropriate bias voltage via the 8-bit DAC,  it will make a 
$\simeq$ 3.9 $\%$ fluctuation in APD gain if $G$=30, as noted 
in $\S$2.2. This fluctuation is not sufficiently small compared to the 
energy resolution of the scintillation detector as described in Table 1. 
Although we are aware 
that a higher-bit DAC (e.g., 16 bit or 32 bit) will improve the 
results further, an 8-bit DAC is the maximum choice onboard Cute-1.7 due to 
limited power consumption and telemetry constraints. In ground based
experiments, the active gain-control system can be more readily refined 
to meet the requirements for various experimental goals.

\section{Conclusion}

We have developed an active gain-control system that stabilizes 
the APD gain under the moderate temperature variations. 
As a performance demonstration, we showed that a scintillation photon 
detector consisting of a reverse-type APD optically 
coupled with a CsI(Tl) crystal, was successfully controlled 
under the temperature variations between $-$20$^{\circ}$C and  
0$^{\circ}$C cycles, and +20$^{\circ}$C and 0$^{\circ}$C cycles. In both 
measurements, the temperature was varied in a 6000 sec cycle 
that mimics the orbital period of future space satellite missions.
The best FWHM energy resolution of 6.1$\pm$0.2 $\%$ was obtained for 
662 keV $\gamma$-rays, with an energy threshold as low as 6.5 keV 
between +20$^{\circ}$C and 0$^{\circ}$C, where the corresponding values 
between $-$20$^{\circ}$C and 0$^{\circ}$C were 6.9$\pm$0.2 $\%$ and 5.2
keV, respectively. These results are comparable with, or only slightly worse 
than, those obtained at fixed temperatures. Our results suggest 
a new potential for APDs with uses in various field of space research and
nuclear physics. For example, we are planning to use the APDs as a 
scientific instrument onboard the $NeXT$ and Cute-1.7 satellite 
missions.  

\section*{Acknowledgements}
We greatly appreciate an anonymous referee for his/her 
comments to improve this manuscript. We also thank Dr. Phil Edwards for 
his constructive suggestions and careful comments to complete this work.
J.Kataoka acknowledges a support by JSPS.KAKENHI (16206080).

\end{document}